\documentclass[12pt,final]{iopart}
\usepackage{graphicx}
\usepackage{rotate}

\begin{document}

\title{Comparing community structure identification}

\author{
  Leon Danon\dag\ddag,\  
  Albert D\'{i}az-Guilera,\dag\  
  Jordi Duch\ddag,\ and
  Alex Arenas\ddag 
} 
  
\address{\dag\ Departament de Fisica
  Fonamental,Universitat de Barcelona, Marti i Franques 1 08086
  Barcelona, Spain} 
\address{\ddag\ Departament d'Enginyeria
  Inform\`{a}tica i Matem\`{a}tiques, Universitat Rovira i Virgili,
  Campus Sescelades, 43007 Tarragona, Spain}
\ead{\tt leon.danon@urv.net}

\begin{abstract}
We compare recent approaches to community structure identification in
terms of sensitivity and computational cost. The recently proposed
modularity measure is revisited and the performance of the methods as
applied to {\em ad hoc} networks with known community structure, is
compared. We find that the most accurate methods tend to be more
computationally expensive, and that both aspects need to be considered
when choosing a method for practical purposes. The work is intended as
an introduction as well as a proposal for a standard benchmark test of
community detection methods.
\end{abstract}

\section{Introduction}

The study of complex networks has received an enormous amount of
attention from the scientific community in recent years
\cite{BARev,NRev,DMRev,Strogatz01,BookBorn,Sitges}. Physicists in
particular have become interested in the study of networks describing
the topologies of a wide variety of systems, such as the world wide
web, social and communication networks, biochemical networks and many
more.  An important open problem is the analysis of modular structure
found in many networks \cite{Newman04}. Distinct modules or
communities within networks can loosely be defined as subsets of nodes
which are more densely linked, when compared to the rest of the
network. Such communities have been observed in different kinds of
networks, most notably in social networks, but also in networks of
other origin such as metabolic or economic networks
\cite{Thurner04,Ravasz02,Guimera05,Holme03}. As a result, the problem of
identification of communities has been the focus of many recent
efforts.

Community detection in large networks is potentially very
useful. Nodes belonging to a tight-knit community are more than likely
to have other properties in common. For instance, in the world wide
web, community analysis has uncovered thematic clusters
\cite{Flake02,Eckmann02}. In biochemical or neural networks,
communities may be functional groups \cite{Zhou05}, and separating the
network into such groups could simplify functional analysis
considerably.

The problem of community detection is quite challenging and has been
the subject of discussion in various disciplines. A simpler version of
this problem, the graph bi-partitioning problem (GBP) has been the
topic of study in the realm of computer science for decades. Here, one
looks to separate the graph into two densely connected communities of
equal size, which are connected with the minimum number of links. This
is an NP complete problem\footnote{In computational complexity theory,
NP (`Non-deterministic Polynomial time') is the set of decision
problems solvable in polynomial time on a non-deterministic Turing
machine. NP-complete problems are the most difficult problems in NP.}
\cite{Garey79}, however several methods have been proposed to reduce
the complexity of the task
\cite{KernighanLin,Fiedler73,Boettcher01a,Pothen90}. In real complex
networks we often have no idea how many communities we wish to
discover, but in general it is more than two. This makes the process
all the more costly. What is more, communities may also be
hierarchical, that is communities may be further divided into
sub-communities and so on
\cite{Guimera03b,Gleiser03,Arenas03,Newman04a}.

Nevertheless, many attempts to tackle these problems have been
proposed recently. The proposed methods vary considerably in terms of
approach and application, which makes them difficult to
compare. Community identification is potentially very useful and
researchers from a number of fields may be interested in using one or
several of the methods for their own purposes. But which? In order for
the reader to be able to make an informed decision as to which method
is most appropriate for which purpose, we distil information from the
literature and compare the performance of those methods which lend
themselves to objective comparison.

To this end, this paper is organised as follows. In section 2
we revisit the modularity measure designed to evaluate how good a
particular partition of a network is. Then, we describe how to measure
the sensitivity of the various methods and suggest the use of a more
accurate representation of algorithm sensitivity based on information
theory. We then compare the methods from a computational cost
perspective and compare their sensitivity when applied to {\it ad hoc}
networks with community structure. Finally, we suggest appropriate
choices of community identification methods for a few different
problems.

\section{Evaluating community identification}
\label{Q}

A question that has been raised in recent years is how a given
partition of a network into communities can be evaluated. A simple
approach that has become widely accepted was proposed in \cite{NG}. It
is based on the intuitive idea that random networks do not exhibit
community structure. Let us imagine that we have an arbitrary network
and an arbitrary partition of that network into $n_c$ communities. It
is then possible to define a $n_c \times n_c$ size matrix ${\mathbf
e}$ where the elements $e_{ij}$ represent the fraction of total links
starting at a node in partition $i$ and ending at a node in partition
$j$.  Then, the sum of any row (or column) of ${\mathbf e}$, $a_i
= \sum_j e_{ij}$ corresponds to the fraction of links connected to
$i$.

If the network does not exhibit community structure, or if the
partitions are allocated without any regard to the underlying
structure, the expected value of the fraction of links within
partitions can be estimated. It is simply the probability that a link
begins at a node in $i$, $a_i$, multiplied by the fraction of links
that end at a node in $i$, $a_i$. So the expected number of
intra-community links is just $a_ia_i$. On the other hand we know that
the {\it real} fraction of links exclusively within a partition is
$e_{ii}$. So, we can compare the two directly and sum over all the
partitions in the graph.

\begin{equation}
Q\equiv\sum_i(e_{ii} - a_i^2)
\end{equation}

This is a measure known as {\it modularity}. As an example, let us
consider a network comprised of $n_c$ fully connected components with
no links between them. If we then have $n_c$ partitions, corresponding
exactly to the components, modularity will have a value of
$1-1/n_c$. As $n_c$ gets large, this value tends to $1$. On the other
hand, for particularly ``bad'' partitions, for example, when all the
nodes are in a community of their own, the value of modularity can
take negative values. This is due to the fact that when nodes are
alone in partitions there can be no internal links.  To avoid this
issue, Massen \& Doye propose an alternative measure \cite{Massen04}.

It is tempting to think that random networks exhibit very small values
of modularity. As Guimer\`{a} {\it et al.} show, this is not the case
\cite{Guimera04}. It is possible to find a partition which not only
has a nonzero value of modularity for random networks of finite size,
but that this value is quite high, for example a network of $128$
nodes and $1024$ links has a maximum modularity of 0.208. This
suggests that these networks that cannot have a modular structure
actually appear to have one due to fluctuations.

\section{Comparative evaluation}
\label{comparison}

The methods that have been presented recently are extremely varied,
and are based on a range of different ideas. In a longer article, we
describe the methods in more detail and classify them according to the
type of approach they present \cite{Danon05book}. Also, the full
description of each can be found in the respective references. Here we
concentrate on comparing the methods in terms of performance. In order
for the reader to be able to compare the algorithms, both in terms of
their speed and sensitivity, we would like to present a qualitative
comparison for all the methods presented until now.  However, this is
not possible as they are very varied, both conceptually and in their
applications.

\begin{figure}
\centerline{\includegraphics*[width=0.7\columnwidth]{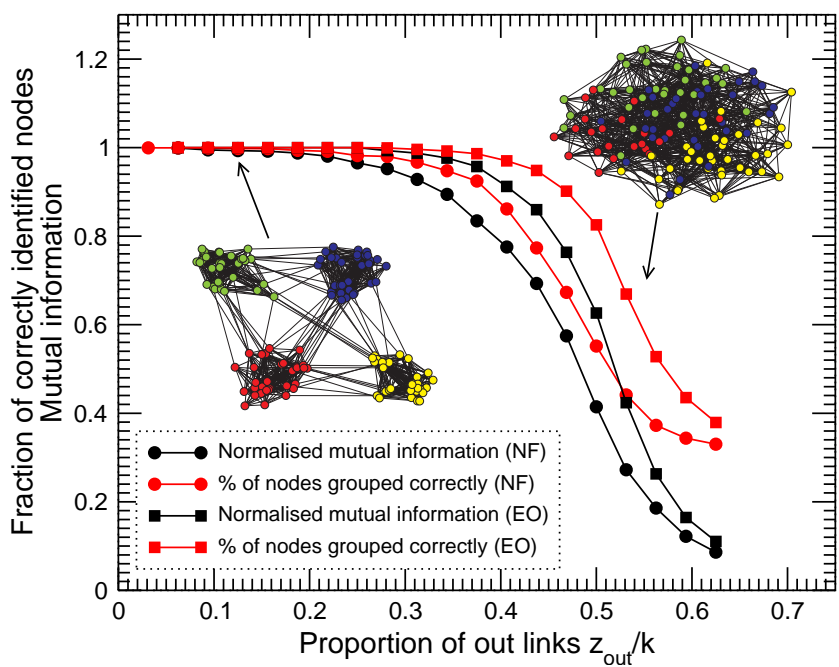}}
\caption{Algorithm sensitivity as applied to ad hoc networks with $n =
 128$, the network divided into four communities of $32$ nodes each
 and total average degree $z_{out}$ fixed to $16$. For low $z_{out}/k$
 the communities are easily distinguished. For higher $z_{out}/k$ this
 becomes more complicated. Both measures of comparing original
 communities to ones found by the detection method are shown. The
 normalised mutual information measure is more discriminatory and
 appears more sensitive to errors in the community identification
 procedure. The results are shown for Newman's fast algorithm
 \cite{Newman04a} and the extremal optimisation algorithm
 \cite{Duch05}.}
\label{fig_ad_hoc}
\end{figure}

One way that has been employed to test sensitivity in many cases is to
see how well a particular method performs when applied to {\it ad hoc}
networks with a well known, fixed community structure \cite{NG}. Such
networks are typically generated with $n = 128$ nodes, split into four
communities containing 32 nodes each. Pairs of nodes belonging to the
same community are linked with probability $p_{in}$ whereas pairs
belonging to different communities are joined with probability
$p_{out}$. The value of $p_{out}$ is taken so that the average number
of links a node has to members of any other community, $z_{out}$, can
be controlled. While $p_{out}$ (and therefore $z_{out}$) is varied
freely, the value of $p_{in}$ is chosen to keep the total average node
degree, $k$ constant, and set to 16. As $z_{out}$ is increased
from zero, the communities become more and more diffuse and harder to
identify, (Figure \ref{fig_ad_hoc}). Since the ``real'' community
structure is well known in this case, it is possible to measure the
number of nodes correctly classified by the method of community
identification.

In \cite{Newman04a}, the author describes a method to calculate this
value. The largest group found within each of the four ``real''
communities is considered correctly classified. If more than one
original community is clustered together by the algorithm, all nodes
in that cluster are considered incorrectly classified. For example,
for the case when $z_{out}/k$ is small, if a method finds three
communities, two of which correspond exactly to two original
communities, and a third, which corresponds to the other two clustered
together, this measure would consider half the nodes correctly
classified. As the author notes, this measure is quite harsh, and some
nodes which one may consider to be correctly clustered are not
counted. On the other end of the spectrum, as $z_{out}/k$ becomes
large, and the networks become essentially random networks, this
method rewards the identification of smaller clusters found within
each of the original communities, which could be misleading.

We suggest that a more discriminatory measure is more appropriate, and
propose the use of the {\it normalised mutual information} measure, as
described in \cite{Kuncheva04,Fred03}. It is based on defining a {\it
confusion matrix} $\bf{N}$, where the rows correspond to the ``real''
communities, and the columns correspond to the ``found''
communities. The element of $\bf{N}$, $N_{ij}$ is the number of nodes
in the real community $i$ that appear in the found community $j$. A
measure of similarity between the partitions, based on information
theory, is then:

\begin{equation}
I(A,B)=\frac{-2\sum^{c_A}_{i=1}\sum^{c_B}_{j=1}
N_{ij}\log\left(\frac{N_{ij}N}{N_{i.}N_{.j}}\right)}
{\sum^{c_A}_{i=1}N_{i.}\log\left(\frac{N_{i.}}{N}\right)
 + \sum^{c_B}_{j=1}N_{.j}\log\left(\frac{N_{.j}}{N}\right)}
\end{equation}

where the number of real communities is denoted $c_A$ and the number
of found communities is denoted $c_B$, the sum over row $i$ of matrix
$N_{ij}$ is denoted $N_{i.}$ and the sum over column $j$ is denoted
$N_{.j}$

If the found partitions are identical to the real communities, then
$I(A,B)$ takes its maximum value of 1. If the partition found by the
algorithm is totally independent of the real partition, for example
when the entire network is found to be one community, $I(A,B)= 0$.

Both measures of accuracy give a good idea of how a method
performs. However, the measure we propose for use here is more
representative of sensitivity if the performance is dubious, since it
measures the amount of information correctly extracted by the
algorithm explicitly. As an example, for small $z_{out}$, where two
original communities are clustered together by the algorithm, this
measure does not punish the algorithm as severely, taking into account
the ability to extract at least some information about the community
structure. On the other hand, for large $z_{out}$, this method is able
to detect that the clusters found by the algorithm have little to do
with the original communities, and $I(A,B) \rightarrow 0$.

\begin{table}
  \centering

\begin{tabular}{|c|c|c|c|}


  \hline

  Author &Ref. & Label & Order \\

  \hline

  \hline

  Eckmann \& Moses&\cite{Eckmann02}&  EM & $O(m\langle k^2\rangle)$ \\

  Zhou \& Lipowsky &\cite{Zhou05} & ZL & $O(n^3)$ \\

  Latapy \& Pons & \cite{Latapy04} & LP & $O(n^3)$ \\

  Newman &\cite{Newman04a} & NF & $O(n\log^2n)$ \\

  Newman \& Girvan &\cite{NG} &  NG & $O(m^2n)$ \\

  Girvan \& Newman &\cite{GN} &  GN & $O(n^2m)$ \\

  Guimer\`{a} et al. & \cite{Guimera04,Guimera05b} & SA & parameter dependent\\

  Duch \& Arenas &\cite{Duch05} & DA & $O(n^2\log n)$ \\

  Fortunato et al. &\cite{Fortunato04} & FLM & $O(n^4)$ \\

  Radicchi et al. &\cite{Radicchi04} & RCCLP & $O(n^2)$ \\

  Donetti \& Mu\~noz&\cite{Donetti04,Donetti05} & DM/DMN & $O(n^3)$ \\

  Bagrow \& Bollt &\cite{Bagrow04}&  BB & $O(n^3)$ \\

  Capocci et al. &\cite{Capocci04}& CSCC & $O(n^2)$ \\

  Wu \& Huberman &\cite{Wu03}& WH & $O(n+m)$ \\

  Palla et al. & \cite{Palla05} & PK & $O(\exp(n))$\\

  Reichardt \& Bornholdt &\cite{Reichardt04} & RB & parameter dependent\\

  \hline

\end{tabular}

\caption{Table summarising how the computational cost of different
approaches scales with number of nodes $n$, number of links $m$ and
average degree $\langle k \rangle$ \cite{Dijkstra}. The labels shown
here are used in Figures \ref{fig_compare} and \ref{678}.}
\label{Table_Orders}

\end{table}

\begin{figure}[!h]
 \centerline{\includegraphics*[width=0.9\columnwidth]{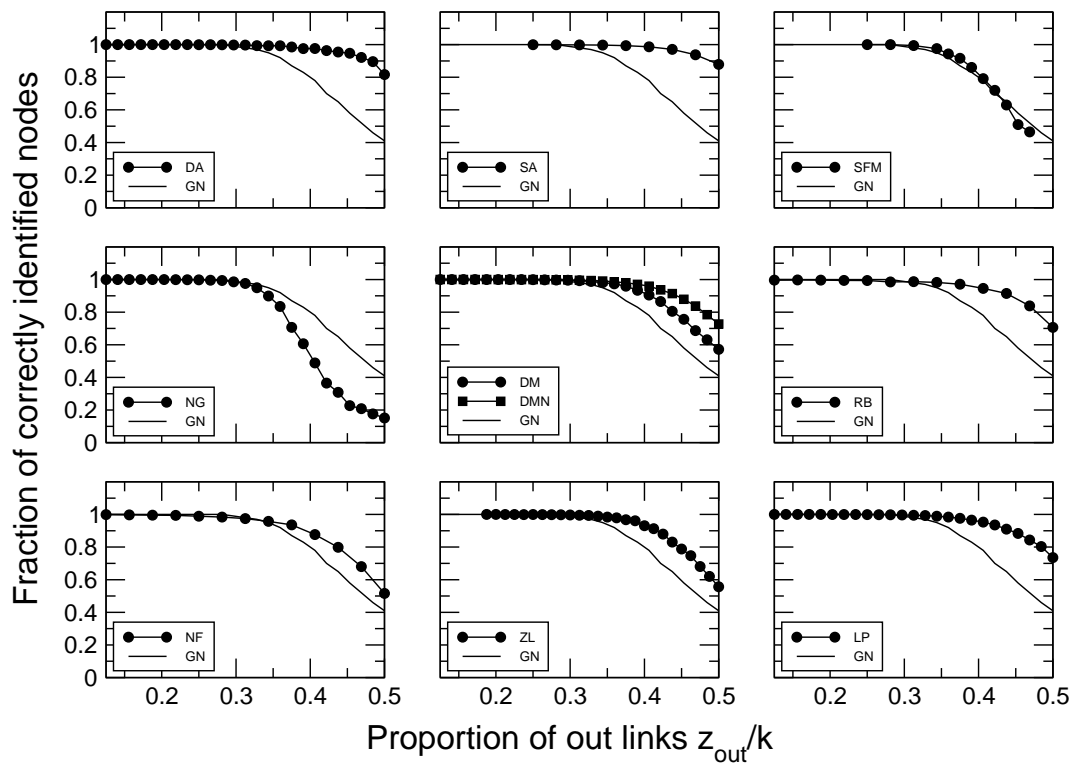}}
\caption{Comparing algorithm sensitivity using ad hoc networks with
 predetermined community structure. The $x$-axis is the proportion of
 connections to outside communities $z_{out}/k$ and the
 $y$-axis is the fraction of nodes correctly identified by the method
 measure as described in \cite{Newman04a}. The labels here correspond
 to the different methods and are listed in Table \ref{Table_Orders}.}
\label{fig_compare}
\end{figure}

\begin{figure}
\centerline{\includegraphics*[width=0.74\columnwidth]{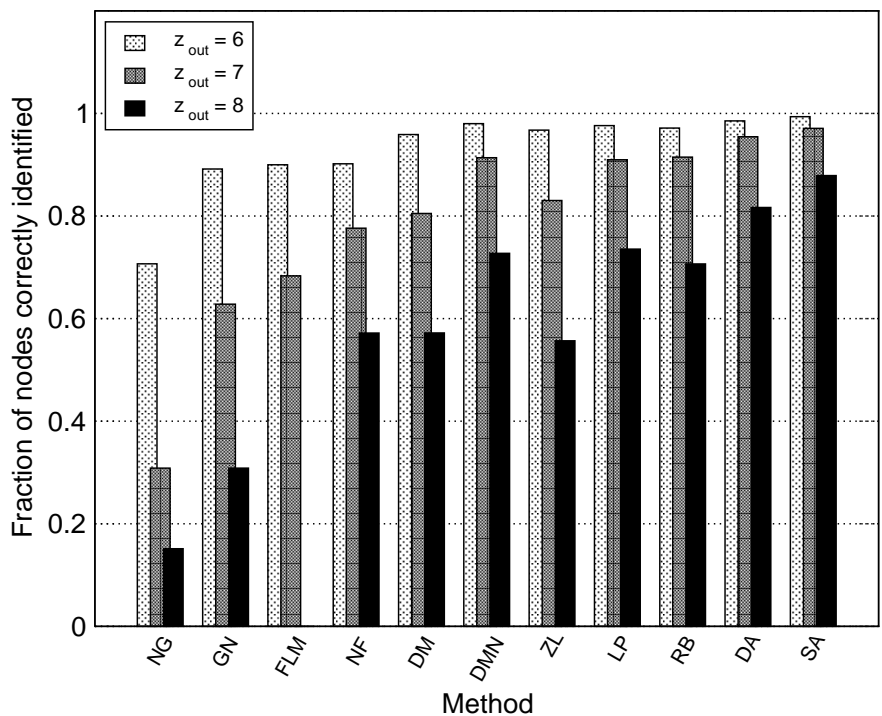}}
\caption{The fraction of correctly identified nodes at three specific
  values of $z_{out}$, $6$, $7$ and $8$ for all available methods and
  for networks with fixed $k=16$. Note that for the FLM method,
  the data for $z_{out}=8$ were not available. Here we can see that
  most of the methods are very good at finding the ``correct''
  community structure for values of $z_{out}$ up to $6$. At $z_{out} =
  7$ some methods begin to falter but most still identify more than
  half of the nodes correctly. At $z_{out} = 8$, when on average half
  the links are external, two methods are still able to identify over 80
  \% of the nodes correctly.}
\label{678}
\end{figure}

In Figure \ref{fig_compare} we show the sensitivity of all methods we
have been able to gather. The percentage of correctly identified nodes
is calculated using the method described in \cite{Newman04a}, since
this is the method employed by the various authors. We can see that
accuracy varies in a similar way across the different methods as
$z_{out}$ increases and the communities become more diffuse. So, it
remains difficult to compare the performance by looking at the methods
separately, even with a reference performance. 

To summarise the large amount of information, in Figure \ref{678} we
plot the fraction of correctly identified nodes for only three values
of $z_{out}$ (6, 7 and 8), corresponding to $z_{out}/k = $ 0.375,
0.4375 and 0.5 respectively, for each method. From this we can see
that most of the methods perform very well for $z_{out}=6$
($z_{out}/k=0.375$), and even for $z_{out}=7$ ($z_{out}/k=0.4375$)
most can identify more than half the nodes correctly. For $z_{out}=8$
($z_{out}/k=0.5$) two methods are still able to identify more than 80
$\%$ of the nodes correctly\footnote{One might expect that as the
proportion of out links approaches $0.5$ community structure no longer
exist. However since the external links are distributed among the
other three communities, individual nodes remain more strongly
connected to their own community than to other communities, even at
this high value of $z_{out}/k$.}.

While accuracy is an essential consideration when choosing a method,
it is just as important to consider the computational effort needed to
perform the analysis \cite{Dijkstra}. For some of the approaches
described in the literature, we have collected estimates of how the
cost scales with network observables. For networks with $n$ nodes and
$m$ links, the methods scale between $O(m+n)$ for the fastest, and
$O(\exp(n))$ for the slowest (Table \ref{Table_Orders}). Such
diversity is due to the different approaches taken by the authors. The
faster methods tend to be approximate and less accurate, while the
slower methods have other advantages (see \cite{Danon05book} for a
more detailed discussion). Differences in speed only become important
when dealing with larger networks.

\section{Choosing an algorithm}

One has to take many factors into account when choosing an algorithm
to use. The above comparison ought to give the reader an idea as to
which algorithm is most appropriate for a given problem. In many
cases, a compromise must be reached between accuracy and running time,
especially for larger networks. To clarify this further, here are a
few examples of real networks, and our suggestion for the
appropriate community identification algorithm.

Say we want to analyse a relatively small network, for example the
metabolic network of the worm {\it Caenorhabditis elegans}, which has
453 nodes. Since the network is small, and current desktop computer
technology is reasonably fast, the speed of the algorithm should pose
no restriction, and one is free to chose the slower, more accurate
methods. In this case the Simulated Annealing (SA) method would be the
most appropriate choice, since it gives the most accurate partitions,
especially if the system is allowed to cool slowly (see
\cite{Guimera04,Massen04,Guimera05b} for more details).

Larger networks, with the number of nodes in the order of $10^5$
become intractable with the more accurate methods. For example, when
attempting to study the community structure of the actor collaboration
network with 374511 nodes, we estimate that the SA algorithm would
take a few months of uninterrupted computation. However, a reasonable
implementation of the fast algorithm would be able to perform this
analysis in just a few hours \cite{Clauset04}, making it the
appropriate choice, even if it's accuracy is not the best.

Let us consider an intermediate sized network such as the Pretty Good
Privacy (PGP) web of trust social network \cite{Guardiola02},
containing 10680 nodes. Although the SA algorithm would run in a
reasonable time, it may be a better choice to compromise and employ a
faster running algorithm. The EO method is not quite as accurate as
SA, but the saving in computational effort for a network of this size
is considerable. It is more accurate than the fast algorithm however,
and so would make it a better choice.

\section{Conclusion}

In this work we have given a brief overview and comparison of the
modern approaches to community identification in complex networks. A
large amount of knowledge has been collected in the field, and real
progress has been made, both in the identification of communities and
their characterisation. Some questions do remain open, and it is these
that we would suggest for further study. Despite these efforts, the
cost involved in computing communities in complex network remains
significant. The fastest algorithm runs in linear time, but this
particular method needs a priori knowledge of the number of expected
communities, and assumes that all communities are of similar size
\cite{Wu03}. At present, the fastest method for finding an unknown
number of communities of unknown sizes has a cost which scales as
$O(n\log^2n)$ with network size. While this makes the analysis of
extremely large networks feasible, this algorithm does not guarantee
that the partition found is the best possible one. Other algorithms
which are more computationally expensive have other merits, such as
accuracy or the ability to identify overlapping communities. So, when
choosing a method one must consider carefully the context of its
use. Ideally, one would like to have a method which guarantees
accuracy and is fast at the same time, but finding such a method is
challenging. The search for faster and more accurate methods is an
important one and we would suggest this for further study.

\ack The authors are grateful to Luca Donetti, Haijun Zhou, Mark
Newman, Santo Fortunato, J\"org Reichardt, Claudio Castellano,
Matthieu Latapy, Jean-Pierre Eckmann and Roger Guimer\`{a} for providing
their data and Sam Seaver for useful comments. This work has been
supported by DGES of the Spanish Government Grant No. BFM-2003-08258
and EC-FET Open Project No. IST-2001-33555. LD gratefully acknowledges
the funding of Generalitat de Catalunya.

\end{document}